\documentclass{aa}
\usepackage{psfig}
\usepackage{amsmath}
\usepackage{natbib}
\newcommand{\home}        {{$_\char126$}}
\newcommand{\mla}         {\langle}
\newcommand{\mra}         {\rangle}
\newcommand{\Rp}          {{\it p}}
\newcommand{\Lp}          {{\it q}}
\newcommand{\RR}          {{pp}}

\newcommand{\EirNot}      {{E^\Rp_{i,\circ}}}
\newcommand{\EilNot}      {{E^\Lp_{i,\circ}}}

\newcommand{\CongjEjlNot} {{E^{\Lp\star}_{j,\circ}}}
\newcommand{\CongjEjrNot} {{E^{\Rp\star}_{j,\circ}}}
\newcommand{\Rgigj}       {{g^\Rp_i g_j^{\Rp^\star}}}
\newcommand{\Laiaj}       {{\alpha^\Lp_i \alpha_j^{\Lp^\star}}}

\newcommand{\sumj}        {\sum\limits_{j \atop {j \ne i}}}

\newcommand{\wijRR}       {w_{ij}^{\Rp\Rp}}

\newcommand{\XijRR}       {X_{ij}^{\RR}}
\newcommand{\giR}         {{g_i^\Rp}}

\newcommand{\aiL}         {{\alpha_i^\Lp}}

\newcommand{\gjR}         {{g_j^\Rp}}

\newcommand{\ajL}         {{\alpha_j^\Lp}}

\newcommand{\SelfCal}     {{self-calibration}}
\newcommand{\antsol}      {{\tt antsol}}
\newcommand{\coPolar}     {{\it co-polar}}
\newcommand{\crossPolar}  {{\it cross-polar}}
\newcommand{\LeakyAntsol} {{\tt leaky~\antsol}}
\newcommand{\rhoRRo}      {{\rho_{ij,\circ}^{\Rp\Rp}}}
\newcommand{\rhoRR}       {{\rho_{ij}^{\Rp\Rp}}}
\newcommand{\rhoLLo}      {{\rho_{ij,\circ}^{\Lp\Lp}}}

\titlerunning{Measurement of polarization leakage}
\authorrunning{Bhatnagar \& Nityananda}

\begin{document}
\bibliographystyle{apj}

\title{Solving for closure errors due to polarization leakage in radio
interferometry of unpolarized sources}

\author{Sanjay Bhatnagar \and Rajaram Nityananda} 

\offprints{Sanjay Bhatnagar}
\institute{National Centre for Radio Astrophysics(TIFR), Post Bag 3,
Ganeshkhind, Pune 411 007, India\\
\email{sanjay@ncra.tifr.res.in, rajaram@ncra.tifr.res.in}
} 

\date{}

\abstract{Mechanical and electronic imperfections can result into
polarization leakage in individual antennas of a radio interferometer.
Such leakages manifest themselves as closure errors even in \coPolar\
visibility measurements of unpolarized sources.  This paper describes
and tests a method for the study of polarization leakage for
radio interferometric telescopes using {\it only} the nominally
\coPolar\ visibilities for unpolarized calibrators.
Interpretation of the resulting closure phases on the Poincar\'e
sphere is presented.  Since unpolarized sources are used, the actual
solutions for leakage parameters is subject to a degeneracy which is
discussed.  This however, does not affect the correction of closure
errors in our scheme. } \maketitle

\section{Introduction}

The mutual coherence function (also called the visibility function)
for an unresolved and unpolarized source, measured by an
interferometer array can be modeled as a product of antenna based
complex gains.  These complex gains can be derived from the measured
visibility function using the standard algorithm, which we call
\antsol. \antsol\ forms the central engine of most amplitude and phase
calibration schemes used for radio interferometric data.  (The
earliest published reference for an algorithm for \antsol\ of which we
are aware is \citet{FREQ_RESPONSE_OF_INTERFEROMETER}).

Usually one measures the components of the incident radiation along
two orthogonal polarization states by using two separate feeds.  The
signals from the two feeds travel through essentially independent
paths till the correlator.
However, due to mechanical imperfections in the feed or imperfections
in the electronics, the two signals can leak into each other at
various points in the signal chain.

At the correlator, signals from all the antennas are multiplied with
each other and the results averaged to produce the visibilities.  The
signals of same polarization are multiplied to produce the \coPolar\
visibilities while the signals of orthogonal polarizations are
multiplied to produce the \crossPolar\ visibilities.  The \coPolar\
and \crossPolar\ visibilities can be used to compute the full Stokes
visibility function.  Antenna based instrumental polarization and
polarization leakage can be derived from the full Stokes coherence
function for a source of known structure (usually an unresolved
source) \cite[henceforth HBS]{HBS1,HBS2}.

The correlator used for the Giant Metrewave Radio Telescope (GMRT) by
default computes the \coPolar\ visibilities using the {\it Indian
mode} of the VLBA Multiplier and Accumulator (MAC) chip.  Here we
describe a method, which we call \LeakyAntsol, for the computation of
the leakages using {\it only} the \coPolar\ visibility function
for an unpolarized source.  Following the notation used by HBS, we
label the two orthogonal polarizations by $\Rp$ and $\Lp$ to remind us
that the formulation is independent of the precise orthogonal pair of
polarization states chosen.

Sect.~\ref{MOTIVATION} describes the motivation which led to this
analysis.  For orientation, Sect.~\ref{ANTSOL} starts with the
problem of solving for the usual complex antenna based gains and sets
up an iterative scheme for the solution.  The problem of {\it
simultaneously} solving for the complex antenna gains and leakages is
then posed in Sect.~\ref{LEAKYANTSOL} and a similar iterative scheme
is set up.  Sect.~\ref{SIMULATIONS} presents the results of the
simulations done to test the scheme.  Sect.~\ref{TEST:CHISQ}
presents some results using the GMRT at 150 MHz.  Also, we were
fortunate to have the L-band feeds of one of the GMRT antennas
converted from linear to circular polarization.  We observed {\tt
3C147} in this mode where all baselines with this special antenna
measured the correlation between nominally linear and circular
polarization.  Results of this experiment demonstrate that the leakage
solutions are indeed giving information about the polarization
properties of the feeds.  These results and their interpretation on
the Poincar\'e sphere are presented in section~\ref{TEST:RX}.
Sect.~\ref{INTERPRET:PS} gives the interpretation of the leakage
solutions and discusses closure errors due to polarization leakage
using the Poincar\'e sphere.

\section{Motivation}
\label{MOTIVATION}

\citet{Rogers_1983} pointed out in the context of the VLBA, that
the polarization leakage cause closure errors even in nominally
\coPolar\ visibilities.  \citet{Massi_1997} have carried out a
detailed study of this effect for the telescopes of the European VLBI
Network (EVN).  Our motivation in this paper is that the current
single sideband GMRT correlator uses the so called {\it Indian mode}
of the VLBA MAC chips to produce only the \coPolar\ visibilities.
Also, the planned Walsh switching has not yet been implemented at the
GMRT and in any case, would not eliminate leakage generated before the
switching point.  Tests done using strong point source dominated
fields show unaccounted closure errors at a few percent level.  The
motivation behind developing an algorithm to solve for gains and
leakages simultaneously, using {\it only} the \coPolar\ visibilities
was to determine if the measured closure errors could be due to
polarization leakage in the system.  Estimates of leakage can then be
used in the primary calibration to remove the effects of polarization
leakage.  This is where this paper differs from the earlier work of
HBS which is about the calibration using the full Stokes visibility
function, needed for observations of polarized sources.  The
polarization leakage in some of the EVN antennas corrupts the
\coPolar\ visibilities at a level visible as a reduction in the
dynamic range of the maps
\citep{Massi_TM_75,Massi_TM_77,Massi_TM_85}.  Thus such a method can
also be used in imaging data from the EVN and other telescopes
affected by such closure errors.

Let $\giR$ represent the complex gain for the $\Rp$-polarization
channel of the $i^{th}$ antenna and $\aiL$ represent the leakage of
the \Lp-polarization signal into the \Rp-polarization channel.  The
electric field measured by antenna $i$ can then be written as

\begin{equation}
E^\Rp_i=g^\Rp_i \EirNot + \aiL \EilNot
\label{EFIELDS}
\end{equation}

where $\EirNot$ and $\EilNot$ are the responses of an {\it ideal}
antenna to the incident radiation in the $\Rp$- and $\Lp$-polarization
states respectively\footnote{Note that Eq.~\ref{EFIELDS} is
equivalent to the Eq.~1 of \citet{Massi_1997}, who use a
different parameterization.  Strictly speaking, the ratio $\aiL/\giR$
is the correct measure of the leakage and this is what has been
plotted in Fig.\ref{FIG:C03_CSQ} below.  This does not however affect
the computational scheme described here.}.  For an unpolarized source
of radiation, $\mla \EirNot \CongjEjlNot \mra = 0$.  The \coPolar\
visibility for such a source, measured by an interferometer using two
antennas denoted by the subscripts $i$ and $j$, is given by

\begin{equation}
\begin{split}
\rhoRR 
       &= \giR \gjR^\star \rhoRRo + \aiL \ajL^\star \rhoLLo + \epsilon_{ij}\\
\label{RHOOBS}
\end{split}
\end{equation}

where $\epsilon_{ij}$ is independent gaussian random baseline based
noise and $\rhoRRo=\mla \EirNot \CongjEjrNot \mra$ and $\rhoLLo=\mla
\EilNot \CongjEjlNot \mra$ are the two ideal \coPolar\ visibilities.
$\epsilon_{ij}$ usually represents the contribution to $\rhoRR$ which
cannot be separated into antenna based quantities.  $\epsilon_{ij}$
therefore is a measure of the intrinsic closure errors in the system
and is usually small.

For an unpolarized point source $\mla \EirNot \CongjEjrNot \mra = \mla
\EilNot \CongjEjlNot \mra = \rhoRRo = I/2$ where $I$ is the total flux
density.  Writing $\XijRR=\rhoRR/\rhoRRo$ we get

\begin{equation}
\label{XIJRR}
\XijRR = \giR \gjR^\star + \aiL \ajL^\star + \epsilon_{ij}
\end{equation}

where $\epsilon_{ij}$ now refers to the baseline based noise in
$\XijRR$.

Assuming $\aiL$s to be negligible, the usual \antsol\ algorithm
estimates $\giR$s such that $\sum_{{i,j}\atop{i \ne j}} \left| \XijRR
- \giR \gjR^\star \right|^2$ is minimized (see section~\ref{ANTSOL}).
Normally, Walsh switching \citep{THOMPSON_AND_MORAN} is used to
eliminate the polarization leakage due to cross-talk between the
signal paths, such that $\aiL\ajL^\star \ll \epsilon_{ij}$.  However,
$\aiL$s can also be finite due to mechanical imperfections in the feed
or the \crossPolar\ primary beam, which cannot be eliminated by Walsh
switching.

In the case of significant antenna based polarization leakage
(compared to $\sqrt{\epsilon_{ij}}$), the second term in
Eq.~\ref{XIJRR} involving $\aiL$s will combine with the closure
noise $\epsilon_{ij}$.  The polarization leakage therefore manifests
itself as increased closure errors (see Sect.~\ref{INTERPRET:PS} for
a geometric explanation on the Poincar\'e sphere).

\section{Algorithm and simulation}
\label{ANTSOL}

In the absence of any polarization leakage, $g_i$s can be estimated by
minimizing

\begin{equation}
S = \sum_{{i,j} \atop {i \ne j}}{\left|\XijRR - \giR\gjR^\star\right|}^2~\wijRR
\label{CHISQ}
\end{equation}

with respect to $g_i$s, where $\wijRR=1/\sigma^2_{ij}$, $\sigma_{ij}$
being the variance on the measurement of $\XijRR$.

In Eq.~\ref{RHOOBS}, if $\rhoRRo$ accurately represents the
source structure, $\XijRR$ will have no source structure dependent
terms and is purely a product of two antenna dependent complex gains.
For a resolved source, $\rhoRRo$ can be estimated from the image of
the source.

Evaluating $\frac{\partial S}{\partial \giR^\star}$ and equating it to
zero\footnote{Complex derivatives can be evaluated by treating $\giR$
and $\giR^\star$ as independent variables
\citep{COMPLEX_ANALYSIS}.}\citep{RANTSOL}, we get



\begin{equation}
\giR~=~{\sum\limits_{j \atop {j \ne i}}~\XijRR \gjR \wijRR \over
\sum\limits_{j \atop {j \ne i}}~\left|\gjR\right|^2 \wijRR}
\label{GI}
\end{equation}

This can also be derived by equating the partial derivatives of $S$
with respect to real and imaginary parts of $\giR^\star$.

Since the antenna dependent complex gains also appear on the
right-hand side of Eq.~\ref{GI}, it has to be solved iteratively
starting with some initial guess for $g_j$s or initializing them all
to 1.  Eq.~\ref{GI} can be written in the iterative form as:

\begin{equation}
\giR^{,n}~=~\giR^{,n-1} + \lambda\left[\giR^{,n-1}-{\sum\limits_{j \atop {j \ne
i}}~\XijRR \gjR^{,n-1} \wijRR\over \sum\limits_{j \atop {j \ne
i}}~\left|\gjR^{,n-1}\right|^2 \wijRR}\right]
\label{GIter}
\end{equation}

where $n$ is the iteration number and $0~<~\lambda~<~1$.  Convergence
would be defined by the constraint $\left|S_n-S_{n-1}\right| < \beta$
(the change in $S$ from one iteration to another) where, $\beta$ is
the tolerance limit and must be related to the average value of
$\epsilon_{ij}$.  Eq.~\ref{GIter} forms the central engine for
the classical \antsol\ algorithm used for primary calibration of the
visibilities and in \SelfCal\ for imaging purposes.  This algorithm
was suggested by \citet{FREQ_RESPONSE_OF_INTERFEROMETER}.

\subsection{The \LeakyAntsol}
\label{LEAKYANTSOL}

In the presence of significant polarization leakage,
Eq.~\ref{XIJRR} can be used to re-write Eq.~\ref{CHISQ} as

\begin{equation}
S = \sum_{{i,j} \atop {i \ne j}} {\left| \XijRR-( \Rgigj + \Laiaj
) \right|}^2~\wijRR
\end{equation}

In this form, $S$ is an estimator for the {\it true} closure
noise $\epsilon_{ij}$ rather than the artificially increased closure
noise ($\aiL \ajL^\star+\epsilon_{ij}$) due to the presence of
polarization leakage.

Equating the partial derivatives $\frac{\partial S}{\partial
{\giR}^\star}$, $\frac{\partial S}{\partial {\aiL}^\star}$ to
zero, we get

\begin{equation}
\giR= { \sumj \XijRR \gjR \wijRR - \aiL \sumj {\ajL}^\star \gjR \wijRR
\over \sumj \left| \gjR \right|^2 \wijRR}
\label{LGI}
\end{equation}

\begin{equation}
\aiL= { \sumj \XijRR \ajL \wijRR - \giR \sumj {\gjR}^\star \ajL \wijRR
\over \sumj \left| \ajL \right|^2 \wijRR}
\label{LAI}
\end{equation}

These non-linear equations can also be iteratively solved.

Eq.~\ref{XIJRR}, which expresses the observed visibilities on a
point source unpolarized calibrator in terms of the gains and leakage
coefficients of the antennas, would take the same form if written in
an arbitrary orthogonal basis. It is clear that the $g$'s and the
$\alpha$'s will change when we change the basis, so this means that
the equations cannot have a unique solution. This situation is
familiar from ordinary self-calibration, when only relative phases of
antennas are determinate, with one antenna acting as an arbitrary
reference.  For observations of unpolarized sources, we can similarly
say that any feed can be chosen as a reference polarization, with zero
leakage, and other feeds have gains and leakages in the basis defined
by this reference. Other conventions may be more convenient, as
discussed in the appendix which discusses degeneracy in detail.

%
%
%

\subsection{Results of the simulations}
\label{SIMULATIONS}

\begin{figure}[h]
\begin{center}
\psfig{file=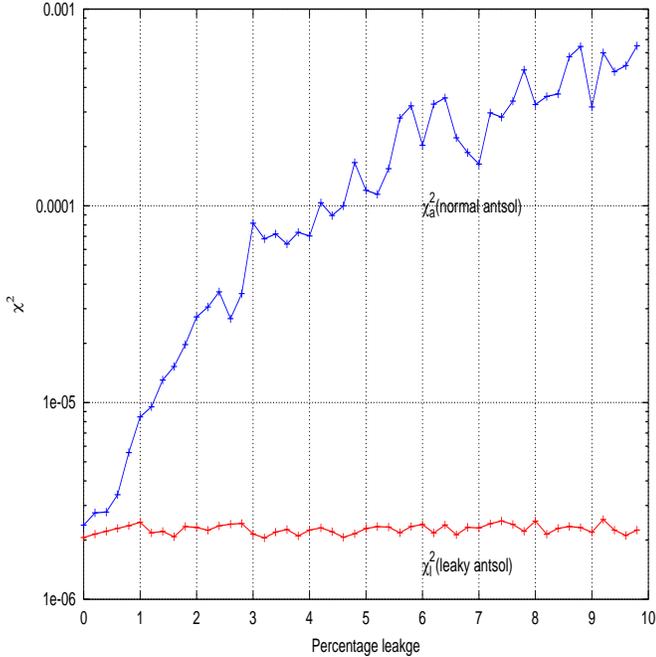,width=3.5truein,height=3.5truein}
\caption[]{Figure showing the results of the simulations.  The top
curve is the value of $\chi^2$ using the classical \antsol\
($\chi^2_\mathrm{a}$).  The bottom curve is the value of $\chi^2$
using the \LeakyAntsol\ ($\chi^2_\mathrm{l}$) as a function of the
percentage polarization leakage.}
\label{FIG:SIM1}
\end{center}
\end{figure}

We simulated visibilities with varying fraction of polarization
leakage in the antennas to test the algorithm as follows.  The antenna
based signal and leakage were constructed as $g_i~=~R_g$ and
$\alpha_i~=f\cdot R_\alpha$ where $R_g$ and $R_\alpha$ were drawn from
the same gaussian random population.  The visibility from two antennas
$i$ and $j$ was then constructed as $X_{ij} = g_i g_j^\star + \alpha_i
\alpha_j^\star + \epsilon_{ij}$ for $0 \le f < 0.1$.  This is
equivalent to a visibility of an unpolarized point source of unit
strength with a complex antenna based gain $g_i$ and leakage
$\alpha_i$ of strength proportional to $f$.  Eq.~\ref{GIter} was
then used to compute $g_i$ and residual $\chi^2$ computed as
$\chi^2_\mathrm{a} = \sum_{ij} \left| 1-\frac{X_{ij}}{g_i
g_j^\star}\right|^2$.  The computed values of $\giR$ were then used to
compute improved estimates for $\giR$ by simultaneously solving for
$\giR$ and $\aiL$ using the iterative forms of Eq.~\ref{LGI} and
\ref{LAI}.  The derived values of $\giR$ and $\aiL$ matched the true
values to within the tolerance limit.  A new $\chi^2$ was computed as
$\chi^2_\mathrm{l} = \sum_{ij} \left| 1-\frac{X_{ij}}{(g_i g_j^\star +
\alpha_i \alpha_j^\star)} \right|^2$.  The values of
$\chi^2_\mathrm{a}$ and $\chi^2_\mathrm{l}$ as a function of $f$ are
plotted in Fig.~\ref{FIG:SIM1}.  The two curves become
distinguishable when the leakage is significantly greater than
$\epsilon_{ij}$ (for $f$ greater than $\sim 1$\%).  After that, the
value of $\chi^2_\mathrm{l}$ is consistently lower than
$\chi^2_\mathrm{a}$, where the contribution of antenna based leakage
has not been removed.  Also notice that $\chi^2_\mathrm{l}$ remains
constant while $\chi^2_\mathrm{a}$ quadratically increases as a
function of $f$.  This is due to the fact that \antsol\ treats the
antenna based polarization leakage as closure errors resulting in an
increased $\chi^2$ with increasing fractional leakage.

\section{Real data}
\subsection{150 MHz data}
\label{TEST:CHISQ}


Engineering measurements for polarization isolation at 150 MHz for the
GMRT show significant polarization leakage in the system.  We
therefore used \LeakyAntsol\ to calibrate the data from the Galactic
plane phase calibrator {\tt 1830-36} which is known to be less than
$0.2\%$ polarized at 1.4 GHz.  The percentage polarization at 150 MHz
is not known, but it is expected to decrease further and it was taken
to be an unpolarized point source.

\begin{figure}[ht]
\begin{center}
\psfig{file=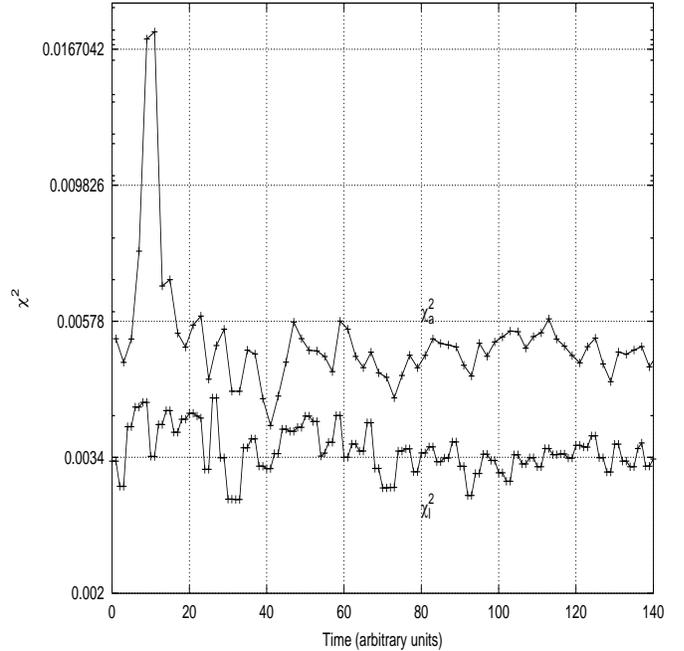,width=3.5truein,height=3.5truein}
\caption[]{Figure showing the results using GMRT data at 150 MHz for
the compact Galactic plane source {\tt 1830-36}.  The top curve is the
value of $\chi^2$ using the classical \antsol\ ($\chi^2_\mathrm{a}$).
The bottom curve is the value of $\chi^2$ using the \LeakyAntsol\
($\chi^2_\mathrm{l}$) as a function of time.}
\label{FIG:RMS@150MHZ}
\end{center}
\end{figure}

Fractional polarization leakage ($\left| \aiL/\giR \right|$) of up to
100\% was measured for most of the antennas, which is consistent with
the estimated leakage measured from system engineering tests.  Again,
$\chi^2_\mathrm{a}$ and $\chi^2_\mathrm{l}$ were computed and the
results are shown in Fig.~\ref{FIG:RMS@150MHZ}.  The 150-MHz GMRT
band suffers from severe radio frequency interference (RFI).  The
sharp rise in the value of $\chi^2_\mathrm{a}$ around sample number 10
is due to one such RFI spike.  This spike is present in the total
power data from all antennas at this time.  On an average, the
$\chi^2$ reduces by $\sim60\%$ when leakage calibration is applied
($\chi^2_\mathrm{l}$).  This is consistent with polarization leakage
being a major source of non-closure at this frequency.

\subsection{L-band data with circular feed}
\label{TEST:RX}

The GMRT L-band feeds are linearly polarized.  For the purpose of a
VLBI experiment conducted in December 2000, the L-band feed of one of
the antennas was converted to a circularly polarized feed.  The rest
of the L-band feeds were linearly polarized and we took this
opportunity to measure correlations between the circularly polarized
antenna with other linearly polarized antennas using the source {\tt
3C147}.  Two scans of approximately one hour long observations were
done using the single side band GMRT correlator.  This correlator
computes only \coPolar\ visibilities.  With this configuration of
feeds, visibilities between the circularly polarized antenna and all
other linearly polarized antennas corresponds to correlation between
the nominal {\tt X-} and {\tt R-}polarizations, labeled by {\tt RX},
were recorded in the first scan.  The polarization of the
circularly polarized antenna was then flipped for the second scan to
record the correlation between the nominal {\tt X-} and {\tt
L-}polarization states, labeled by {\tt LX}.

The VLA Calibrator Manual\footnote{The VLA Calibrator manual is
available on the web from {\tt http://www.aoc.nrao.edu/\home
gtaylor/calib.html}} lists the percentage polarization
($\frac{\sqrt{Q^2 + U^2 + V^2}}{I}$) for {\tt 3C147} at L-band
$<0.1\%$.  The \crossPolar\ terms in Eq.~\ref{RHOOBS}, which are
assumed to be zero, will therefore contribute an error of the order of
$0.1\%$.  These \crossPolar\ terms will be, however, multiplied by
gains of type $\giR\aiL^\star$.  Since $\giR$ and $\aiL$ are both
assumed to be uncorrelated between antennas, this error will manifest
as random noise in Eq.~\ref{XIJRR}.  Within the limits of other
sources of errors, the source {\tt 3C147} can therefore be considered
to be a completely unpolarized source.

\begin{figure}[ht]
\begin{center}
\psfig{file=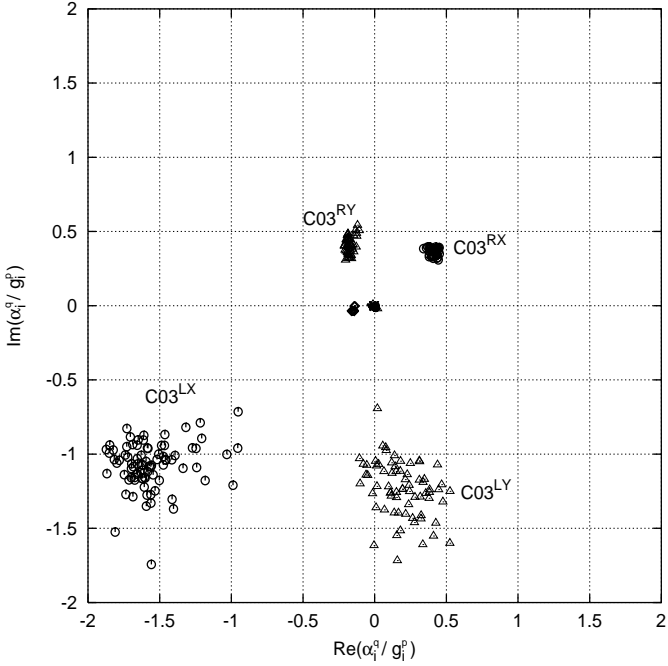,width=88truemm,height=88truemm}
\caption[]{Figure showing the results using visibilities with one
circularly polarized antenna and all other linearly polarized antennas
at L-band.  The x- and y-axis denote the real and imaginary parts of
$\aiL/\giR$ respectively.  $\giR$ and $\aiL$ were solved for every
integration time ($\sim 17 s$).  All linearly polarized antennas are
close to the origin in this plot.  The solutions for the circularly
polarized antenna (C03) are the set of points away from the origin
(shown by open circles and triangles).  The two sets of points for
this antenna, separated from each other by $\sim 180^\circ$ are
solutions for the right- and left-circular polarization channel.  The
points denoted by open circles are from correlation between the right-
and left-circular polarization of C03 with nominal linear
X-polarization of the other antennas (labeled as $\mathrm{C03^{RX}}$
and $\mathrm{C03^{LX}}$ respectively).  The points denoted by
triangles are from correlation of C03 with nominal Y-polarization of
the other antennas (labeled as $\mathrm{C03^{RY}}$ and
$\mathrm{C03^{LY}}$).}
\label{FIG:C03_CSQ}
\end{center}
\end{figure}

\subsubsection{Results and Interpretation}

The response of an ideal circularly polarized antenna to unpolarized
incident radiation can be expressed as a superposition of two linear
polarization states as $E_{i,\circ}^R = E_{i,\circ}^X e^{\iota \delta}
+ E_{i,\circ}^Y e^{-\iota \delta}$ where, the superscripts $R$, $X$
and $Y$ denote the right circular and the two linear polarization
states respectively.
$\delta$ is half the phase difference between the two linear
polarization states and is equal to $\pi/4$ for right-circular
polarization and $-\pi/4$ for left-circular polarization.  Writing the
general Eq.~\ref{EFIELDS} for right-circularly polarized antenna
as $E^R_i=g^R_i E^R_{i,\circ} + \alpha_i^L E^L_{i,\circ}$ and
substituting for $E_{i,\circ}^R$ and $E_{i,\circ}^L$ we get

\begin{equation}
\begin{split}
E^R_i =& g^R_i \left(E^X_{i,\circ} e^{\iota \delta} + E^Y_{i,\circ} e^{-\iota \delta}
\right) +\\
        & \alpha^L_i \left(E^X_{i,\circ} e^{-\iota \delta} + E^Y_{i,\circ} e^{\iota \delta} \right) 
\end{split}
\end{equation}

Eq.~\ref{XIJRR} for the case of correlation between a circularly
polarized and a linearly polarized antenna, with polarization leakage
in both the antennas, can be written as

\begin{equation}
\begin{split}
X^{RX}_{ij} =&  (g_i^R e^{\iota \delta} + \alpha_i^L e^{-\iota
\delta}) g_j^{X^\star} + \\ 
& (g_i^R e^{-\iota \delta} + \alpha_i^L e^{\iota
\delta}) \alpha_j^{Y^\star} \\
 \equiv& g_i^{X^\prime} g_j^{X^\star} + \alpha_i^{Y^\prime} \alpha_j^{Y^\star}
\end{split}
\end{equation}

where $g_i^{X^\prime} = g_i^R e^{\iota \delta} + \alpha_i^L e^{-\iota
\delta}$ and $\alpha_i^{Y^\prime} = g_i^R e^{-\iota \delta} +
\alpha_i^L e^{\iota \delta}$.  The \LeakyAntsol\ solutions for the
circularly polarized antenna in this case will correspond to
$g_i^{X^\prime}$ and $\alpha_i^{Y^\prime}$.

Let $P_i=\alpha_i^Y/g_i^X$ ($P_i=\alpha_i^{Y^\prime}/g_i^{X^\prime}$
for the circularly polarized antenna).  Then, the amplitude of $P_i$
is a measure of the fractional polarization leakage in the antenna
while the phase of $P_i$ gives the phase difference between the signal
from one of the feeds and the leaked signal from the other feed.  For
an ideal circularly polarized antenna, $|P_i|\approx 1$.  A plot of
the real and imaginary parts of this quantity for all antennas should
therefore clearly show $P_i$ for the circularly polarized antenna with
an amplitude of 1 and at an angle of $\pi/2$ with respect to the
nominal X-axis.

The real and imaginary parts of $P_i$ for all antennas from this
experiment are shown in Fig.~\ref{FIG:C03_CSQ}.  The solutions were
computed for every integration cycle of $\sim 17$~sec and the points
on this plot represent the tip of phasor $P_i$.  The collection of
points near the origin are for all the linearly polarized antennas
while the collection of two sets of points away from the origin,
approximately an angle of $\pi$ from each other, are for the
circularly polarized antenna.  The solutions found by \LeakyAntsol\
match the expected results quite well. This therefore constitutes a
reasonably controlled test with real data showing that the solutions
indeed provide information about the polarization leakage in the
system.

This experiment however provides much more information about the
polarization properties of the antenna feeds used.  The collection of
points in the first quadrant denoted by open circles are the values of
$P_i$ derived from the correlation between the nominal
right-circularly polarized signal and the linearly polarized signals
along the nominal X-axis from all other antennas.  Points in the third
quadrant are similarly derived using the left-circular signals.  The
set of points denoted by triangles in the second and fourth quadrant
are derived using correlations of right- and left-circularly polarized
signals with the linearly polarized signals along the nominal Y-axis
from all other antennas.

A larger spread in the solutions using the left-circularly polarized
signals indicates that the closure noise (from other unknown sources)
in these signals is higher.  The fact that the amplitude of $P_i$
derived using the right-circularly polarized signals is $\approx 0.5$
indicates that the nominal circularly polarized feed is in fact
elliptically polarized with this axial ratio.  The spread of $\pm
1-2\%$ about the origin is indicative of polarization leakage at the
level of few percent in the linearly polarized antennas as well.  The
leakage in one of the linearly polarized antennas is significantly
larger ($\approx 4\%$).  Since this kind of data is routinely taken on
primary calibrators during GMRT observations for synthesis imaging,
\LeakyAntsol\ provides a useful diagnostic of system health,
polarization performance and numbers needed to correct the data in
high accuracy work.

The following test was also carried out to check that the closure
phase on a triangle involving the circular feed was indeed mainly due
to polarization effects.  The three baselines making up this triangle
were flagged as bad baselines from the input data and a new solution
found for the gains and leakages of all antennas.  This solution
predicted the same closure phase (to within errors) as actually
observed.


\section{Closure phase and the Poincar\'e sphere}
\label{INTERPRET:PS}

In this section we use right- and left-circular polarization states as
the basis.  A general elliptically polarized state can be written as a
superposition of two states represented by the vector $\left(
\begin{array}{ll} \cos \theta/2 \\ \sin \theta/2~e^{\iota \phi} \\
\end{array} \right)$.  Clearly, $\theta = \pi/2$ corresponds to linear
polarization and $\theta \ne 0,\pi/2$ to elliptical polarization.
Increasing $\phi$ by $\zeta$ rotates the direction of the linear state
or the major axis of the ellipse by $\zeta/2$.  We can chose the phase
of the basis so that $\phi=0$ corresponding to linear polarization
along the x-axis.  The Poincar\'e sphere representation of the state
of polarization maps the general elliptic state to the point ($\theta,
\phi$) on the sphere.  The properties of this representation are
reviewed by \citet{Ramachandran_Ramaseshan}.  We are concerned here
with one remarkable property, discovered by
\citet{PoincareSphere,Panch_Collected_Works}.  Whenever there is
constructive interference between two sources of radiation, it is
natural to regard them as in phase.  An unexpected property of
this simple definition manifests itself when we consider 3 sources of
radiation of different polarization - that if a source A is in phase
with B and B in phase with C, C in general need not be in phase with
A.  The phase difference between A and C is known in the optics
literature as the geometric or Pancharatanam phase (see also
\citet{Ramaseshan_Rajaram, BerrysPhase}).  We show that this naturally
occurs in radio interferometry of an unpolarized source with three
antennas of different polarizations.

Let the polarization states of the three antennas be represented by
$\left( \begin{array}{ll} g_1 \\ \alpha_1 \\ \end{array} \right)$,
$\left( \begin{array}{ll} g_2 \\ \alpha_2 \\ \end{array} \right)$, and
$\left( \begin{array}{ll} g_3 \\ \alpha_3 \\ \end{array} \right)$ in a
circular basis.  Denoting the vector $\left(\begin{array}{ll} g_i \\
\alpha_i \\ \end{array} \right)$ by $\psi_i$, one clearly see that the
visibility on the 1-2 baseline is proportional to $\psi_1^\dag
\psi_2$.  Hence the closure phase around a triangle made by antennas
1, 2, and 3 is the phase of the complex number (also called the triple
product) $V_{123}=(\psi^\dag_1 \psi_2)(\psi^\dag_2 \psi_3)(\psi^\dag_3
\psi_1)$.  In the quantum mechanical literature, this type of quantity
goes by the name of Bargmann's invariant and its connection to the
geometric phase was made clear by \citet{Samuel_Bhandari_1988}.  With
some work, one can give a general proof that the closure phase (phase
of $V_{123}$) is equal to half the solid angle subtended at the centre
of the Poincar\'e sphere by the points represented by $\psi_1$,
$\psi_2$, and $\psi_3$ on the surface of the sphere.  For the case
where the polarization state of the three antennas are same, this
phase is zero in general.  However, when the polarization states of
the antennas are different, this phase is non-zero.

The well known result that an arbitrary polarization state can be
represented as a superposition of two orthogonal polarization states
translates to representing any point on the Poincar\'e sphere by the
superposition of two diametrically opposite states on a great circle
passing through that point.  For example, circular polarization can be
expressed by two linear polarizations, each with intensity
$1/\sqrt{2}$.  In the context of the present work, the nominally
circularly polarized antenna maps to a point away from the equator on
the Poincar\'e sphere (it would be exactly on the pole if it is purely
circular) while the rest of the antennas map close to the equator
(they would be exactly on the equator if they are purely linear and
map to a single point if they were also identical).  The visibility
phase due to the extra baseline based term in Eq.~\ref{XIJRR} due
to polarization mis-match is a consequence of the Pancharatanam phase
mentioned above.  This phase, on a triangle involving the circularly
polarized antenna, will be close to the angle between the two linear
antennas.  For example, if $\psi_1=\left( \begin{array}{ll} 1 \\ \iota
\\ \end{array} \right)$, $\psi_2=\left( \begin{array}{ll} 1 \\ 0 \\
\end{array} \right)$, and $\psi_3=\left(\begin{array}{ll} \cos \gamma
\\ \sin \gamma \\ \end{array} \right)$, the phase of $V_{123}$ will be
$\gamma$.  This picture can be depicted by plotting the real and
imaginary parts of $\aiL/\giR$, which is done in
Fig.~\ref{FIG:C03_CSQ}.  The circularly polarized antenna can be
clearly located in this figure as the set of point away from the
origin while the linearly polarized antennas as the set of points
close to the origin.  The collection of points located away but almost
symmetrically about the origin represents the nominal right- and
left-circularly polarized feeds.  Points on the equator, but
significantly away from the origin represents an imperfect linearly
polarized antenna.  Note that the average closure phase between the
nominally linear antennas is close to zero, which defines the mean
reference frame in Fig.~\ref{FIG:C03_CSQ}.


\section{Conclusions}
\citet{Rogers_1983} pointed out that non ideal feed polarizations of
the individual antennas of a radio interferometer can result into
closure errors in the \coPolar\ visibilities.  In this paper we
describe and demonstrate a method to measure the polarization leakage
of individual antennas using the nominally \coPolar\ visibilities for
an unpolarized calibrator.  This method can therefore be used as a
useful tool for studying the polarization purity of the antennas of
radio interferometers from the observations of unpolarized
calibrators.  However, since only unpolarized calibrators are used,
the actual solution for the leakage parameters is subject to a
degeneracy.  This degeneracy does not affect the correction of the
visibilities and can be used to remove the closure errors due to
polarization leakage.  \citet{Massi_1997} have shown that such
polarization leakage induced closure errors in the data from the EVN
is the dominant effect of instrumental polarization.  For the EVN,
this effect can be seen as a reduction in the dynamic range of the
images.  Our method can be used for such data to remove these closure
errors for unpolarized sources.

The general elliptic state of the polarization of radiation can be
represented by a point on the Poincar\'e sphere.  The phase difference
between three coherent sources of radiation but with different states
of polarization goes by the name of Pancharatanam or geometric phase
in the optics literature.  We interpret the \coPolar\ visibilities
with polarization leakages on the Poincar\'e sphere and show that the
polarization induced closure phase errors in radio interferometers is
same as the Pancharatanam phase of optics.  The antenna based leakages
also map to points on the Poincar\'e sphere and the ambiguity in the
solution can be understood as a rigid rotation of the Poincar\'e
sphere, which leaves the leakage solutions unchanged relative to each
other.

\begin{acknowledgements}

We thank A.~Omar and S.~Roy who suggested that there could be closure
errors in the data due to polarization leakage.  Many useful
discussions on this with A.~P.~Rao and Divya~Oberoi are thankfully
acknowledged.  We also thank the telescope operators and other GMRT
staff for their help and cooperation in making possible the
observations used for this work.  The GMRT is run by the National
Centre for Radio Astrophysics of the Tata Institute of Fundamental
Research.

This research has made use of NASA's Astrophysics Data System Abstract
Service.  All of this work was done using computers running the
GNU/Linux operating system and it is a pleasure to thank all the
numerous contributors to this software.
\end{acknowledgements}

\appendix
\section{Non-uniqueness of solutions}

We discuss the non-uniqueness of the solutions of
Eq.~\ref{XIJRR}, and possible convenient conventions for choosing
a specific solution. One obvious degeneracy is that multiplication of
all the $\alpha$'s by one common phase factor independent of antenna,
and all the $g$'s, by another, in general different, common factor,
does not affect the right hand side of Eq.~\ref{XIJRR}. Also, the
equation was written in a specific basis, say right and left
circular. But it would have had the same form when using any other
orthogonal pair as basis since the source is unpolarized. Hence
we are free to apply this change of basis to one solution to get
another solution of Eq.~\ref{XIJRR}. Under such a change, the
coefficients transform according to

\begin{equation}
\left( 
\begin{array}{ll} 
g^\prime \\ 
\alpha^\prime \\ 
\end{array} 
\right)=
\left( 
\begin{array}{cc} 
\cos\phi                 & e^{\iota\gamma} \sin\phi \\
-e^{-\iota\gamma}\sin\phi & \cos\phi                  \\ 
\end{array} 
\right)
\left( 
\begin{array}{ll} 
e^{\iota\zeta_1}& 0                \\ 
0               & e^{\iota\zeta_2} \\
\end{array} 
\right)
\left( 
\begin{array}{ll} 
g \\ 
\alpha \\ 
\end{array} 
\right)
\end{equation}

It is easy to verify that under this change, $\alpha_i^\prime
\alpha_j^{\star\prime} + g_i^\prime g_j^{\star\prime}=\alpha_i
\alpha_j^\star+g_i g_j^\star$.  Clearly, since $\chi^2$ is unchanged
by these transformations, an iterative algorithm will simply pick one
member of the set of possible solutions, determined by the initial
conditions. Having found one such, one could apply a suitable
transformation to obtain a solution satisfying some desired
condition. For example, if one has nominally linear feeds, one might
impose the statistical condition that there is some mean linear basis
with respect to which the leakage coefficients will be as small as
possible.  Such a condition has the advantage that a perfect set of
feeds is not described in a roundabout way as a set of leaky feeds
with identical coefficients, simply because the basis chosen was
different.  Carrying out the minimization of $\sum\left|g_i\right|^2$
by the method of Lagrange multipliers, subject to a constant $\chi^2$,
we obtain the condition that $\sum\alpha_i^*g_i=0$. This solution can
be interpreted as requiring the leakage coefficients to be orthogonal
to the gains, and is reasonable when we think about the opposite kind
of situation, when the leakages are "parallel" to the gains,
i.e. identical apart from a multiplicative constant.  In such a case,
we would obviously change the basis to make the new leakage zero.  If
we have a solution which does not satisfy this orthogonality
condition, we can bring it about in two steps. First, choose an
overall phase for the $\alpha$'s so that $\sum \alpha_i^\star g_i$ is
real. Then, carry out a rotation in the $g - \alpha$ plane by an angle
$\theta$ satisfying $\tan \theta=\sum \alpha_i^* g_i /(\sum(g_i
g_i^\star -\alpha_i \alpha_i^\star)$. This rotation has been so chosen
that it makes the leakage "orthogonal" to the gains, in the sense
required above.  Even after this is done, we still have the freedom to
define the phase zero independently for the two orthogonal
states. This is because we are only dealing with unpolarized
sources. Of course, if we had a linearly polarized calibrator, the
relative phase of right and left circular signals would not be
arbitrary.

A more geometric view of this degeneracy is obtained when we think in
terms of the Poincar\'e sphere representation of the states of
polarization of all the feeds. The cross correlation between the
outputs of two feeds, both of which receive unpolarized radiation, has
a magnitude equal to the cosine of half the angle between the
representative points on the sphere. Measurements of all such cross
correlations with unpolarized radiation fixes the relative geometry of
the points on the sphere, while leaving a two parameter degeneracy
corresponding to overall rigid rotations of the sphere. This
degeneracy can be lifted by the measurement of one polarized source at
many parallactic angles.

Finally, we note that for the purpose of correcting the observations
of unpolarized sources for the effects of non-identical feed
polarization, the degeneracy is unimportant, because the correction
factor is precisely the right hand side of Eq.~\ref{XIJRR} which
is unaffected by all the transformations we have discussed.

\bibliography{Paper}

\end{document}